%% file: lishep.tex
\documentclass[%
 reprint,
 amsmath,amssymb,
 aps,
]{revtex4-1}

\usepackage{graphicx}
\usepackage{dcolumn}
\usepackage{bm}
\usepackage{epsfig}

\def \et {E_{T}}
\newcommand{\met}{\mbox{$\not\!\!\et$}}
\begin{document}


\title{Diffraction via Pomeron and photon exchanges using proton tagging at the LHC}%

\author{Christophe Royon}
 \email{christophe.royon@cea.fr}
\affiliation{CEA/IRFU/Service de physique des particules, CEA/Saclay, 91191 
Gif-sur-Yvette cedex, France}

\date{\today}

\begin{abstract}
We present some physics topics that can be studied at the LHC using proton
tagging. We distinguish the QCD (Pomeron structure, BFKL
analysis...) from the exploratory physics topics (HIggs boson, anomalous
couplings between photons and $W/Z$ bosons).
\end{abstract}

\maketitle


In this short report, we discuss some potential measurements to be performed
using proton tagging detectors at the LHC. We can distinguish two kinds of
measurements. The first motivation of these detectors is to probe
the structure of the colorless exchanged object, the Pomeron, in terms of 
quarks and gluons, in order to constrain further its structure and to reach
a domain of
energy unexplored until today. We can also mention tests of the Balitsky Fadin
Kuraev Lipatov (BFKL) evolution euqation~\cite{bfkl} using gap between jets
in diffractive events. The second motivation is to explore more rare events
sunch as the production of the Higgs boson and the search for beyond
standard model physics such as quartic anomalous couplings between photons and
$W/Z$ bosons and $\gamma$. We assume in the following intact protons to be tagged 
in dedicated detectors located at about 210 m for ATLAS (220 m for CMS) as
described at the end of this report. In the following, most of the presented results
assume the ATLAS proton detectors to be installed and similar results are expected
for CMS.

\section{Inclusive diffraction measurement at the LHC}
In this section, we discuss potential measurements at the LHC that can constrain
the Pomeron structure, the colorless object which is exchanged in diffractive
interactions. The Pomeron structure in terms of quarks and gluons has been
derived from QCD fits at HERA and at the Tevatron and it is possible to probe this structure and the
QCD evolution at the LHC in a completely new kinematical domain.

\subsection{Jet production in double Pomeron exchanges processes}

The high energy and luminosity at the LHC allow the exploration of a 
completely new kinematical domain. Once can first probe if the Pomeron is universal between
$ep$ and $pp$ colliders, or in other
other words, if we are sensitive to the same object at HERA and the LHC. 
Tagging both diffractive protons in 
ATLAS and CMS will allow the QCD evolution of the gluon and quark densities
in the Pomeron to be tested and compared with the HERA measurements. In addition,
it is possible to measure the gluon and quark densities 
using the dijet and $\gamma + jet$ productions.
The different diagrams of the processes that can be studied at the LHC
are shown in Fig.~1, namely double pomeron exchange (DPE) production of dijets (left),
of $\gamma +$jet (middle), sensitive respectively to the gluon and quark contents of the
Pomeron, and the jet gap jet events (right). 
As an example, we show in 
Fig.2, the ratio of the number of events obtained for $\gamma+$jet and dijet production
for a luminosity of 300 pb$^{-1}$ when the protons are 
tagged in the ATLAS Forward Physics detectors (AFP) at 210 m. The acceptance of the AFP
proton detectors at 210 m from the ATLAS interaction point (resp. 210 and 420 m) in terms
of proton energy loss is $0.015 <\xi<0.15$ (resp. $0.0015 < \xi < 0.15$). The acceptance is
similar for the CMS detector . Fig.~2 shows the potential of this
measurement to constrain further the quark density in the Pomeron. The QCD fits to the proton
diffractive structure functions and to the dijet cross section performed at HERA do not allow to
constrain the differences between the quark densities in the Pomeron, and the quark densities are
assumed to be equal in the fit ($u=\bar{u}=d=\bar{d}=s=\bar{s}$). At the LHC, we will be able to
test this hypothesis by constraining further the quark densities in the Pomeron as shown in
Fig.~2. Let us notice that the best observable is the ratio of the total diffractive mass computed by measuring
the two intact protons in the final state between $\gamma+$jet and dijet events since most
systematics uncertainties cancel in the ratio. Fig.~2 displays the ratio obtained for a luminosity
of 300 pb$^{-1}$ which corresponds to three weeks of data at ,low luminosity at the
LHC~\cite{matthias}.

\subsection{Jet gap jet production in double Pomeron exchanges processes}
This process is illustrated in Fig~1, right~\cite{jgjpap,fwd}. Both protons are intact 
after the interaction and detected in AFP at 210 m, two jets are measured in the 
ATLAS central detector and a gap devoid of any energy is present between 
the two jets. This kind of event is important since it is sensitive to QCD 
resummation dynamics given by the BFKL~\cite{bfkl} 
evolution equation . This process has never been measured to date and will be 
one of the best methods to probe these resummation effects, benefitting from 
the fact that one can perform the measurement for jets separated by a large 
angle (there is no remnants which `pollute' the event). As an example, the 
cross section ratio for events with gaps to events with or without gaps 
as a function of the leading jet $p_T$ is 
shown in Fig~3 for 300 pb$^{-1}$. The measurement has to be performed at medium luminosity
at the LHC so that the gap between the jets is not ``polluted" by pile up events. The presence
of few pile up events in average is still possible for this measurement since central can be 
identified using central tracks fitted to the main vertex of the event. It is worth noticing that
the ratio between the jet gap jet and the dijet cross sections in DPE events is of the order of 20\% which is
much higher than the expectations for non-diffractive events. This is due to the fact that the
survival probability of 0.03 at the LHC does need to be applied for diffarctive events.

\begin{figure}
\begin{center}
\epsfig{file=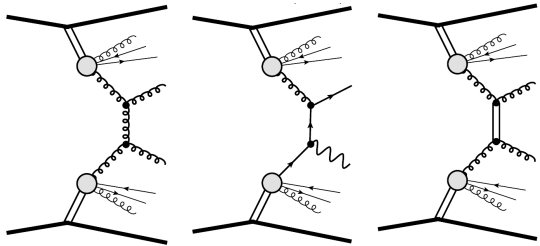,width=7.5cm}
\caption{Inclusive diffractive diagrams. From left to right: jet production in inclusive double
pomeron exchange, $\gamma +$jet production in DPE, jet gap jet events}
\label{d0b}
\end{center}
\end{figure}

\begin{figure}[t]
\hfill
\begin{minipage}[t]{.45\textwidth}

\centerline{\includegraphics[width=1.2\columnwidth]{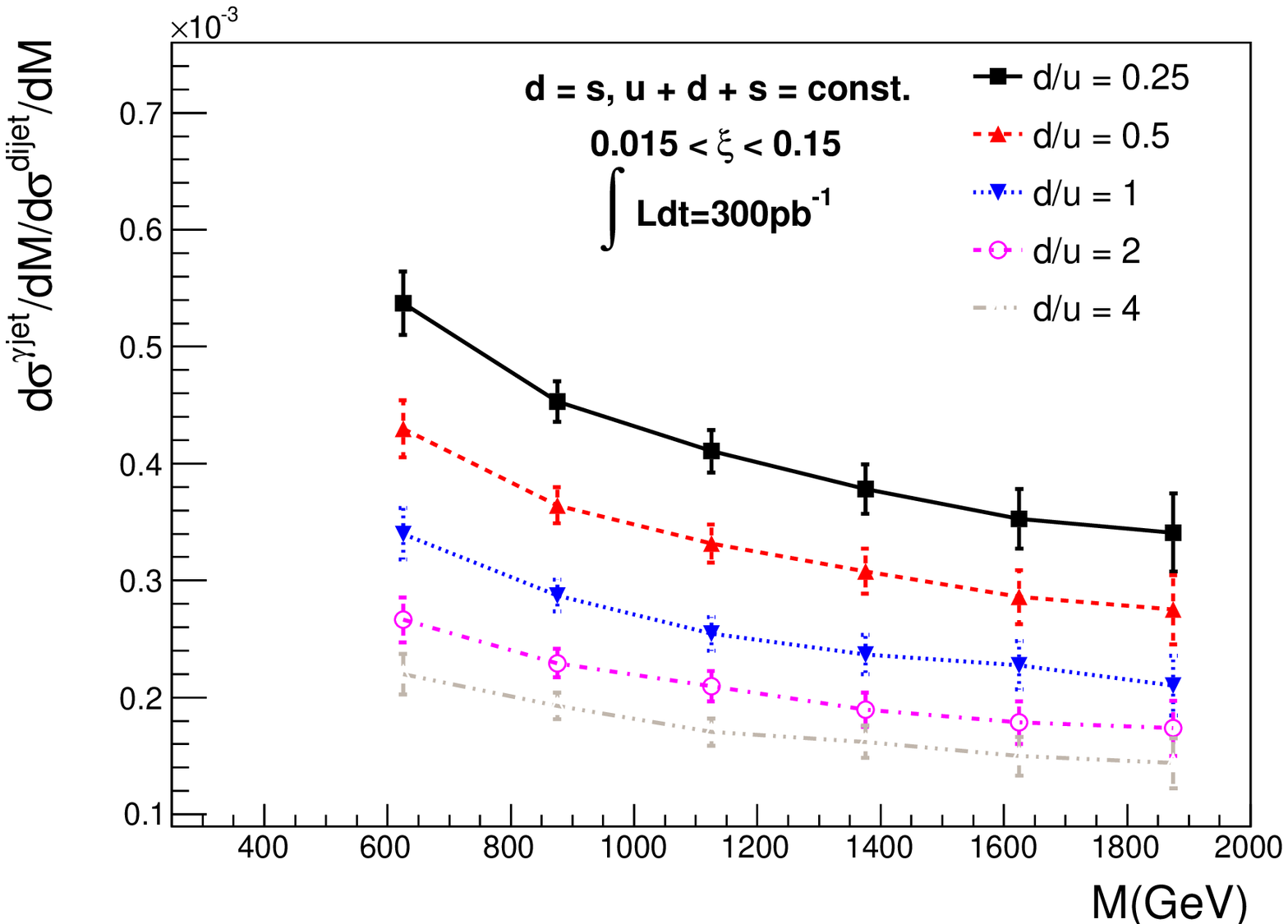}}
\caption{Ratio between the $\gamma+$jet and dijet productions and sensitivity on the quark
structure in the Pomeron.}
\label{uncert1}

\end{minipage}
\hfill
\begin{minipage}[t]{.45\textwidth}

\centerline{\includegraphics[width=1.1\columnwidth]{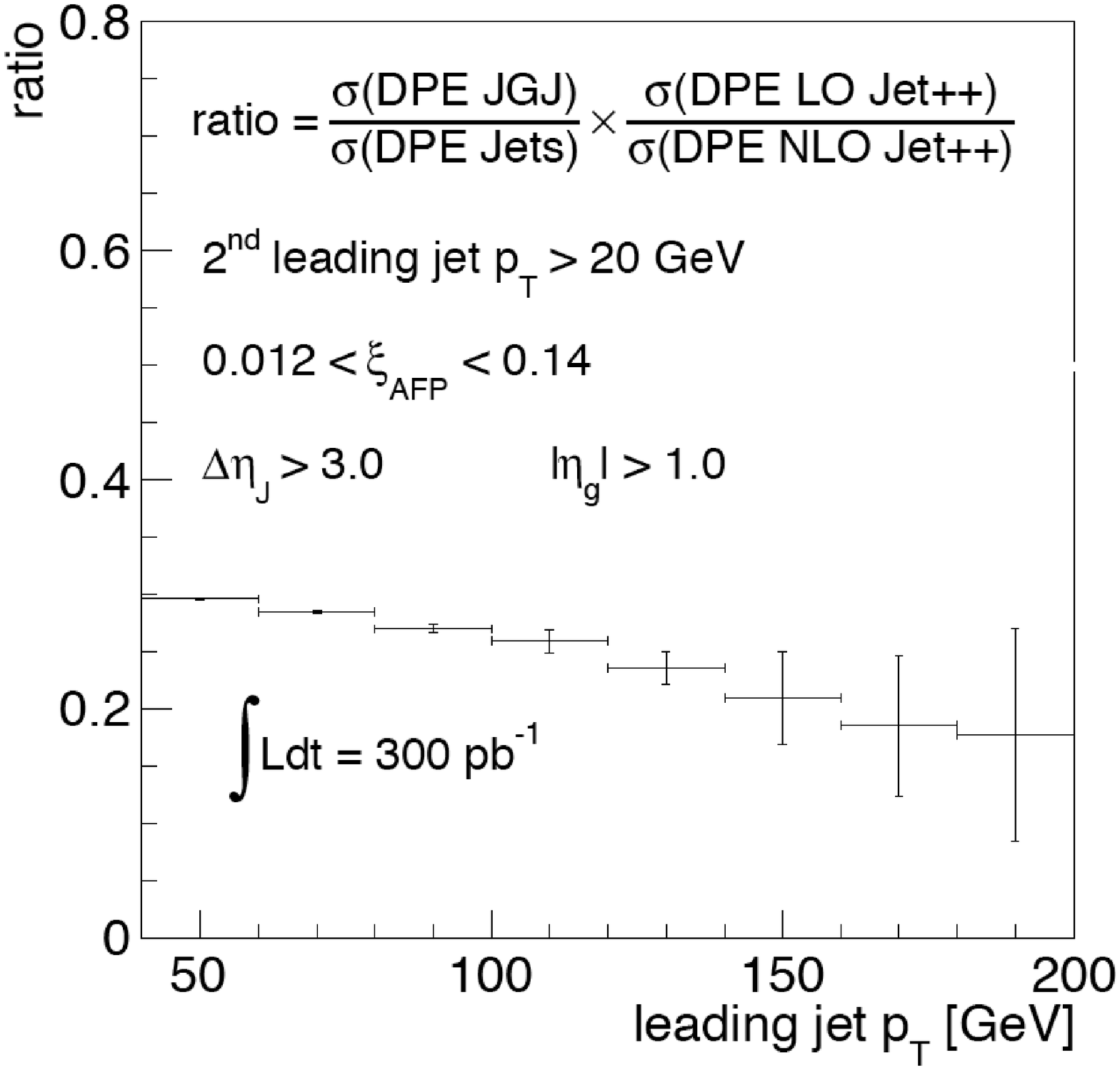}}
\caption{Ratio of DPE Jet gap jet events to standard DPE dijet events as a
function of the leading jet $p_T$~\cite{jgjpap}.}
\label{uncert2b}
\end{minipage}
\hfill
\end{figure}

\section{Exclusive jet and Higgs boson production at the LHC}

The Higgs and jet exclusive production in both Khoze Martin Ryskin~\cite{kmr}
(KMR) and 
CHIDe~\cite{chide} models have been
implemented in the Forward Physics Monte Carlo (FPMC)~\cite{FPMC}, a generator
that has been designed to study forward physics, especially at the LHC. It aims
to provide a variety of diffractive processes in one common framework,
\textit{i.e.} single
diffraction, double pomeron exchange, central exclusive production and
two-photon exchange.

There are three main sources of uncertainties in exclusive models:
the gap survival probability which will be measured using the first
LHC data (in this study we 
assume a value of 0.03 at the LHC for a center-of-mass energy of 14 TeV), 
the unintegated gluon density which appears in the exclusive cross section
calculation and which contains the hard and the
soft part (contrary to the hard part, the soft one is not known precisely and
originates from a phenomenological parametrisation), and finally  
the limits of the Sudakov integral which have not been all fixed by
theoretical calculations~\cite{usb,kmr}.

In order to constrain the uncertainty on the Higgs boson cross section, 
we study the possible constraints using early LHC measurements of
exclusive jets using an integrated luminosity of 100 pb$^{-1}$. In
addition to the statistical uncertainties, we consider a conservative $3\%$ jet
energy scale uncertainty as the dominant contribution to the systematic error.
We obtain the 
prediction for Higgs boson production as shown in Fig.~4 (yellow band). 

In addition, it is possible to measure the exclusive jet cross section 
(see the process in Fig.~6) at high jet $p_T$ at the LHC
benefitting from the high luminosity accumulated. The expectation 
is shown in Fig.~5 for the ATLAS experiment as an example. 
The results of the measurement are shown as
black points for a luminosity of 40 fb$^{-1}$ and 23 pile up events, assuming
the protons to be detected in AFP at 210 m~\cite{loi}.
The expected contributions from background (non-diffractive, single diffractive
with pile up and double pomeron exchange events) are shown as well as the
exclusive jet one in yellow. The statistical significance of the measurement is
up to 19$\sigma$.

\begin{figure}[t]
\hfill
\begin{minipage}[t]{.45\textwidth}

\centerline{\includegraphics[width=1.\columnwidth]{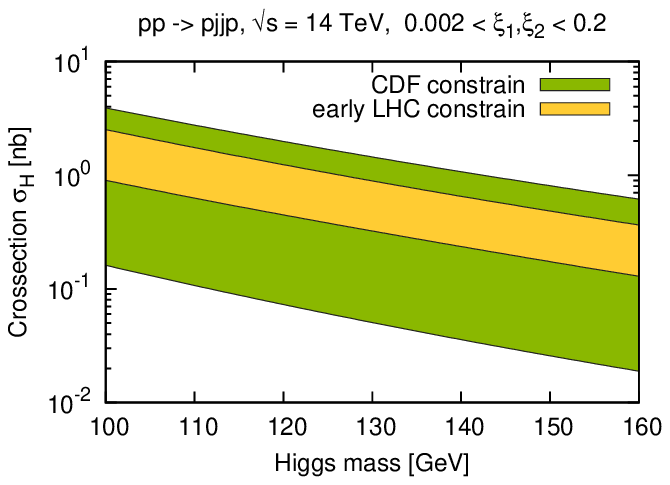}}
\caption{Total uncertainty for exclusive Higgs production at the LHC: 
constraint provided by 
the CDF measurements
and a possible measurement at the LHC with a luminosity of 100 pb$^{-1}$~\cite{usb}.}
\label{uncert1b}

\end{minipage}
\hfill
\begin{minipage}[t]{.45\textwidth}

\centerline{\includegraphics[width=1.1\columnwidth]{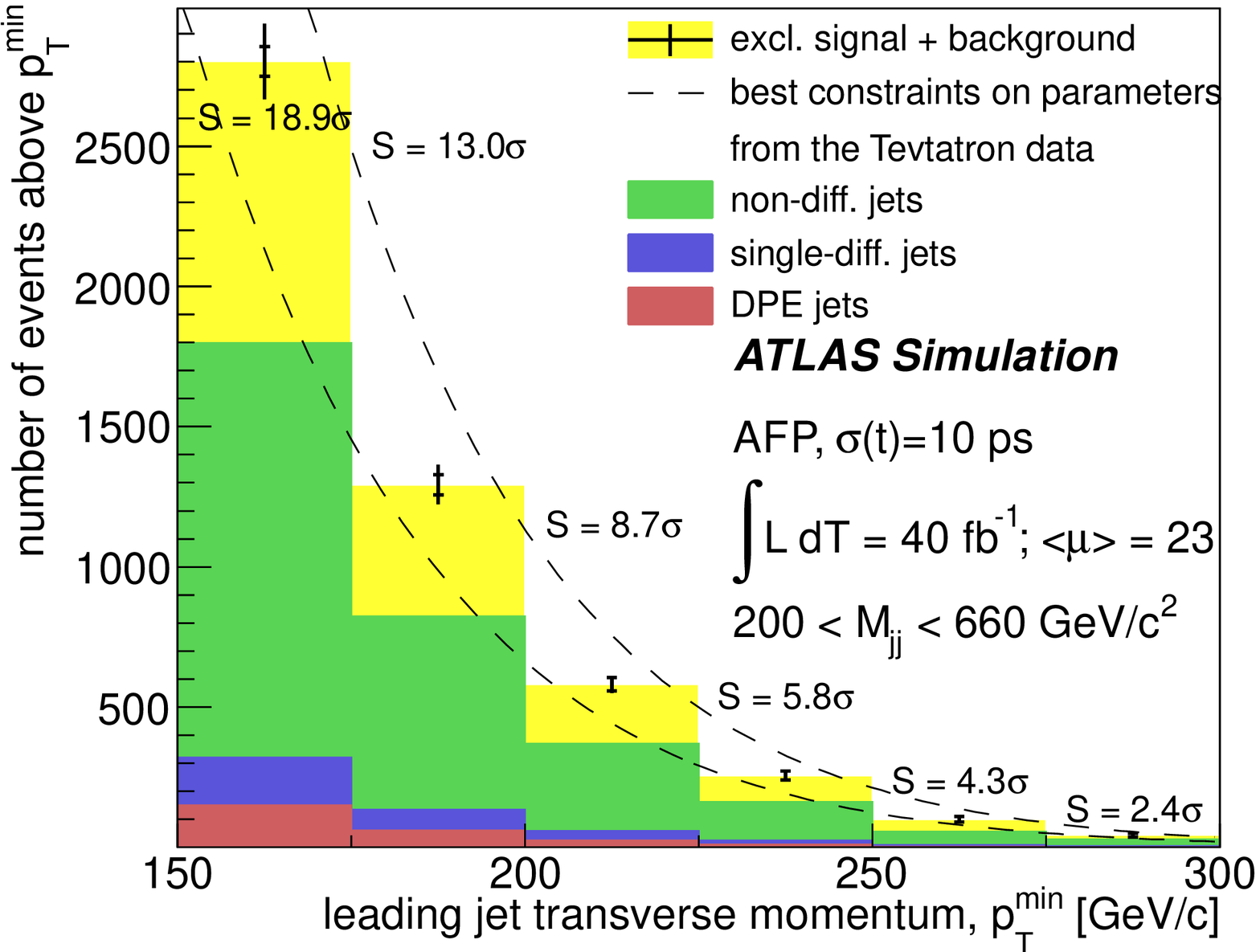}}
\caption{Measurement of the exclusive jet production at high jet $p_T$ at the LHC in the ATLAS
experiment~\cite{loi} .}
\label{uncert2}
\end{minipage}
\hfill
\end{figure}

\section{Exclusive $WW$ and $ZZ$ production}

In the Standard Model (SM) of particle physics, the couplings of fermions and 
gauge bosons are constrained by the gauge symmetries of the Lagrangian.
The measurement of $W$ and $Z$ boson pair productions via the exchange of
two photons  
allows to provide directly stringent tests
of one of the most important and least understood
mechanism in particle physics, namely the
electroweak symmetry breaking.

\subsection{Photon exchange processes in the SM}
The process that we intend to study is the $W$ pair production shown in Fig.~7
induced by the 
exchange of two photons~\cite{us}. It is a pure QED process
in which the decay products of the $W$ bosons are measured in the central 
detector and the scattered protons leave intact in
the beam pipe at very small angles and are detected in AFP.

After simple cuts to select exclusive $W$ pairs decaying into leptons, such
as a cut on the proton momentum loss of the proton ($0.0015<\xi<0.15$) --- we
assume the protons to be tagged in AFP at 210 and 420 m ---
on the transverse momentum of the leading and second leading leptons at 25 and
10 GeV respectively, on $\met>20$ GeV, $\Delta \phi>2.7$ between leading
leptons, and $160<W<500$ GeV, the diffractive mass reconstructed using the
forward detectors, the background is found to be less than 1.7 event for 30
fb$^{-1}$ for a SM signal of 51 events~\cite{us}. 

\begin{figure}[t]
\hfill
\begin{minipage}[t]{.35\textwidth}

\epsfig{file=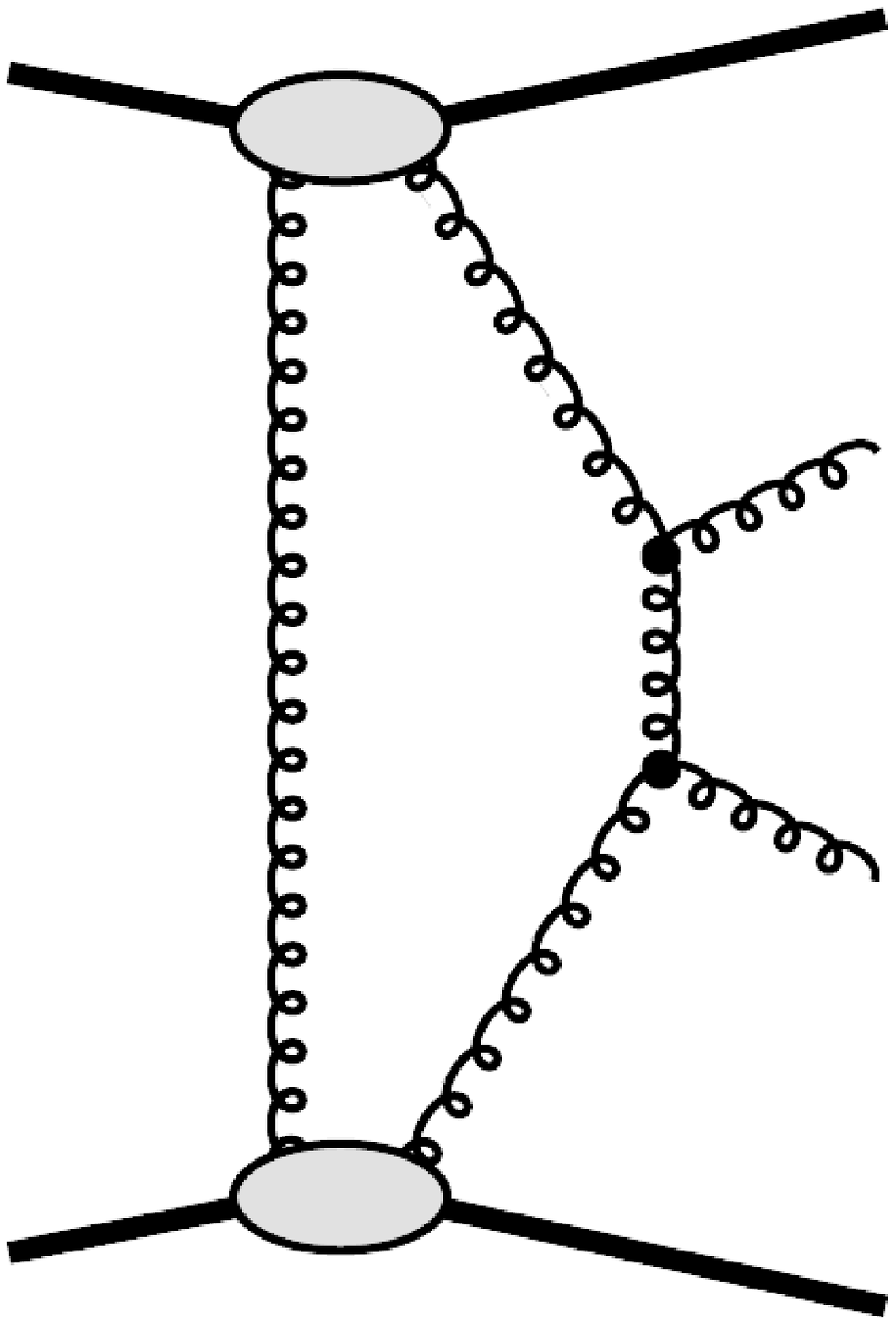,width=3.9cm} 
\caption{Exclusive jet production.}

\end{minipage}
\hfill
\begin{minipage}[t]{.45\textwidth}

\epsfig{file=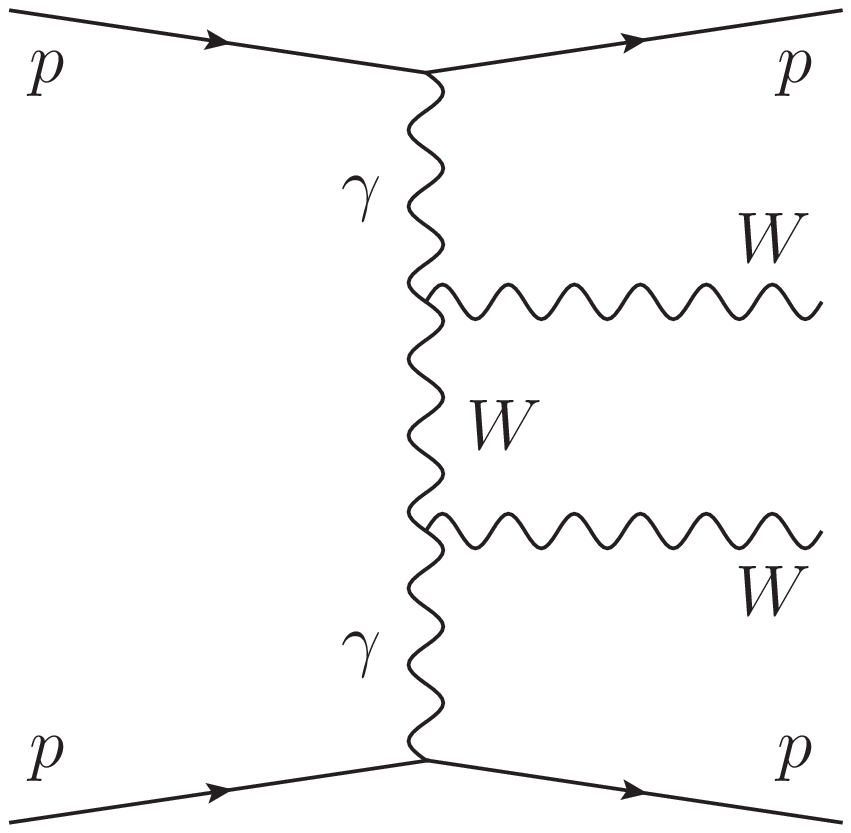,width=5.cm} 
\caption{Diagram showing the two-photon production of $W$ pairs.}

\end{minipage}
\hfill
\end{figure}

\begin{figure}
\begin{center}
\epsfig{file=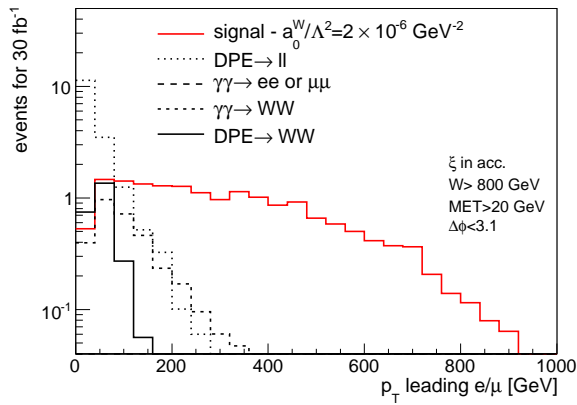,width=8.cm}
\caption{Distribution of the transverse momentum of the leading lepton for
signal and background after the cut on $W$, $\met$, and $\Delta \phi$ between
the two leptons~\cite{us}.}
\label{d0}
\end{center}
\end{figure}

\subsection{Quartic anomalous couplings}
The parameterization of the quartic couplings
based on \cite{Belanger:1992qh} is adopted. 
The cuts to select quartic anomalous gauge coupling $WW$ events are similar as the
ones we mentioned in the previous section, namely $0.0015<\xi<0.15$ for the
tagged protons corresponding to the AFP detector at 210 and 420 m, $\met>$ 20 GeV, 
$\Delta \phi<3.13$ between the two leptons. In
addition, a cut on the $p_T$ of the leading lepton $p_T>160$ GeV and on the
diffractive mass $W>800$ GeV are requested since anomalous coupling events
appear at high mass. Fig~8 displays the $p_T$ distribution of the leading lepton
for signal and the different considered backgrounds. 
After these requirements, we expect about 0.7 background
events for an expected signal of 17 events if the anomalous coupling is about
four orders of magnitude lower than the present LEP limit ($|a_0^W / \Lambda^2| =
5.4$ 10$^{-6}$) for a luminosity of 30 fb$^{-1}$. The strategy to select anomalous coupling $ZZ$ events is
analogous and the presence of three leptons or two like sign leptons are 
requested. 
Table 1 gives the reach on anomalous couplings at the LHC for
luminosities of 30 and 200 fb$^{-1}$ compared to the present OPAL limits~\cite{opal}. We note that we can gain almost
four orders of magnitude in the sensitivity to anomalous quartic gauge couplings
compared to LEP experiments, and it is possible to reach the values expected in 
extra-dimension models. The tagging of the protons using the ATLAS Forward
Physics detectors is the only method at present to test so small values of
quartic anomalous couplings.

\begin{table}
\begin{center}
   \begin{tabular}{|c||c|c|c|}
    \hline
    Couplings & 
    OPAL limits & 
    \multicolumn{2}{c|}{Sensitivity @ $\mathcal{L} = 30$ (200) fb$^{-1}$} \\
    &  \small[GeV$^{-2}$] & 5$\sigma$ & 95\% CL \\ 
    \hline
    $a_0^W/\Lambda^2$ & [-0.020, 0.020] & 5.4 10$^{-6}$ & 2.6 10$^{-6}$\\
                      &                 & (2.7 10$^{-6}$) & (1.4 10$^{-6}$)\\ \hline               
    $a_C^W/\Lambda^2$ & [-0.052, 0.037] & 2.0 10$^{-5}$ & 9.4 10$^{-6}$\\
                      &                 & (9.6 10$^{-6}$) & (5.2 10$^{-6}$)\\ \hline               
    $a_0^Z/\Lambda^2$ & [-0.007, 0.023] & 1.4 10$^{-5}$ & 6.4 10$^{-6}$\\
                      &                 & (5.5 10$^{-6}$) & (2.5 10$^{-6}$)\\ \hline               
    $a_C^Z/\Lambda^2$ & [-0.029, 0.029] & 5.2 10$^{-5}$ & 2.4 10$^{-5}$\\
                      &                 & (2.0 10$^{-5}$) & (9.2 10$^{-6}$)\\ \hline               
    \hline
   \end{tabular}
\end{center}
\caption{Reach on anomalous couplings obtained in $\gamma$ induced processes
after tagging the protons in AFP compared to the present OPAL limits. The $5\sigma$ discovery and 95\%
C.L. limits are given for a luminosity of 30 and 200 fb$^{-1}$~\cite{us}} 
\end{table}

The search for quartic anomalous couplings between $\gamma$ and $W$ bosons was
performed again after a full simulation of the ATLAS detector including pile
up~\cite{loi} assuming the protons to be tagged in AFP at 210 m only. 
Integrated luminosities of 40 and 300 fb$^{-1}$ with, 
respectively, 23 or 46 average pile-up
events per beam crossing have been considered. In order to reduce the
background, each $W$  
is assumed to decay leptonically (note that the semi-leptonic case in under study). 
The full list of background processes 
used for the ATLAS measurement of Standard Model $WW$ cross-section was
simulated, namely $t \bar{t}$, $WW$, $WZ$, $ZZ$, $W+$jets, Drell-Yan and 
single top events. In addition, the additional diffractive backgrounds
mentioned in the previous paragraph were also simulated,
The requirement of the presence of at least one
proton on each side of AFP within a time window of 10 ps allows us to reduce the 
background by
a factor of about 200 (50) for $\mu =$ 23 (46). The $p_T$ of the leading 
lepton originating from the
leptonic decay of the $W$ bosons is required to be 
$p_T >$ 150 GeV, and that of the next-to-leading
lepton $p_T>$ 20 GeV. Additional requirement of the dilepton mass to 
be above 300 GeV allows us to remove
most of the $Z$ and $WW$ events. Since only leptonic decays of the 
W bosons are considered, we require in addition less than 3 tracks associated 
to the primary vertex, which allows us to reject a large fraction of the
non-diffractive backgrounds (e.g. $t \bar{t}$, diboson
productions, $W+$jet, etc.) since they show much higher track multiplicities. 
Remaining Drell-Yan and
QED backgrounds are suppressed by requiring the difference in azimuthal angle between the
two leptons $\Delta \phi <$ 3.1.  After these requirements, a similar
sensitivity with respect to fast simulation without pile up was obtained. 

Of special interest will be also the search for anomalous quartic $\gamma \gamma \gamma
\gamma$ anomalous couplings which is now being implemented in the FPMC
generator. Let us notice that there is no present existing limit on such
coupling and the sensitivity using the forward proton detectors is expected to
be similar as the one for $\gamma \gamma WW$ or $\gamma \gamma ZZ$ anomalous
couplings. If discovered at the LHC, $\gamma \gamma \gamma
\gamma$ quartic anomalous couplings might be related to the existence of
extra-dimensions in the universe, which might lead to a reinterpretation of
some experiments in atomic physics. As an example, the Aspect photon correlation
experiments~\cite{aspect} might be interpreted via the existence of 
extra-dimensions. Photons
could communicate through extra-dimensions and the deterministic interpretation
of Einstein for these experiments might be true if such anomalous
couplings exist. From the point of view of atomic physics, the results of the
Aspect experiments would depend on the distance of the two photon sources.

\section{Forward Proton Detectors in ATLAS and CMS}

In this section, we describe the proposal to install the ATLAS Forward 
Proton (AFP)
detector in order to detect intact  protons at 206 and 214 meters on both side of
the ATLAS experiment~\cite{loi} (similar detectors will be installed around CMS
by TOTEM/CMS). 
This one arm will consist of two sections (AFP1 and AFP2) contained in a special design of
beampipe or in more traditional roman pots. In the first section (AFP1), 
a tracking station composed by 6 layers of Silicon detectors will be deployed.
The second station AFP2 will contain 
another tracking station similar to the one already described and a timing
detector. In addition, a similar structure could be installed at about 420 m from the ATLAS
interaction point.
The aim of this setup, mirrored by an identical arm placed on the opposite side
of the ATLAS interaction point,
will be to tag the protons emerging intact from the $pp$ interactions so allowing ATLAS to exploit
the program of diffractive and photoproduction processes described in the previous sections. 

\subsection{Movable beam pipes}
The idea of movable Hamburg beam pipes is quite simple~\cite{HBP}: a larger section of the 
LHC beam pipe than the usual one
can move close to the beam using bellows so that the detectors located at its edge (called pocket)
can move close
to the beam by about 2.5 cm when the beam is stable (during injection, the detectors are in
parking position). 
In its design, the 
predominant aspect is the minimization of the thickness of the portions 
called floor and window (see Fig.~ \ref{Figa}). Minimizing the depth of the floor ensures that the
detector can go as close to the beam as possible allowing us to detect protons scattered at very small
angles, while minimizing the depth of the thin window is important to keep the protons intact and
to reduce the impact of multiple interactions. 
Two configurations exist for the movable beam pipes: the first one at 206 m form the ATLAS interaction
point hosts a Si detector 
(floor length of about 100 mm) and the second one (floor 
length of about 400 mm) the timing and the Si detectors. While it is still being discussed if the
movable beam pipe or the roman pot solution will be chosen at 210 m, it is clear that movable beam
pipes are favoured at 420 m since not enough space is available at this position and
new cryostats have been developped to host these movable beam pipes in the cold region at 420 m.
The usage of roman pots at 420 m would require a costly cryogenic bypass to be installed to isolate
the region where roman pots would be installed.

\begin{figure}
\centering
\includegraphics[width=3.5in]{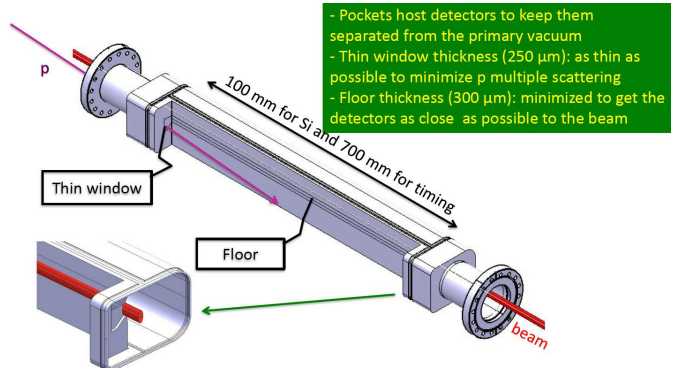}
\caption{Scheme of the movable beam pipe.}\label{Figa}
\end{figure}

\subsection{3D Silicon detectors}
The purpose of the tracker system is to measure points along the trajectory of 
beam protons that are deflected at small angles as a result of collisions. 
The tracker when combined with the LHC dipole and quadrupole magnets,
forms a powerful momentum spectrometer. Silicon tracker stations will be
installed in  Hamburg beam pipes or roman pots at $\pm$ 206 and $\pm$ 214 m from the 
ATLAS interaction point (and also at 420 m later if these additional detectors are appoved). 

The key requirements for the silicon  tracking system at 220~m are:
\begin{itemize}
\item Spatial resolution of $\sim$ 10 (30) $\mu$m per detector station in $x$ ($y$)
\item Angular resolution for a pair of detectors of about 1~$\mu$rad
\item High efficiency over the area of 20~mm $\times$ 20~mm corresponding to the distribution of
diffracted protons
\item Minimal dead space at the edge of the sensors allowing us to measure the scattered protons at
low angles
\item Sufficient radiation hardness in order to sustain the radiation at high luminosity
\item Capable of robust and reliable operation at high LHC luminosity 
\end{itemize}

The basic building unit of the AFP detection system is a module consisting of 
an assembly of a sensor array, on-sensor read-out chip(s), electrical services,
data acquisition and detector control system. The module will be
mounted on the mechanical support with embedded cooling and other necessary
services. The sensors are double sided 3D 50$\times$250 micron pixel detectors with slim-edge 
dicing built by FBK and CNM companies. The sensor efficiency has been measured to be close to 100\% 
over the full size in beam tests. A possible upgrade of this device will be to
use 3D edgeless Silicon
detectors built in a collaboration between SLAC, Manchester, Oslo, Bergen...
A new 
front-end chip FE-I4 has been developed for the Si detector by the Insertable B Layer (IBL)
collaboration in ATLAS. The FE-I4 integrated circuit contains 
readout circuitry for 26 880 hybrid pixels arranged in 80~columns on 250~$\mu$m
pitch by 336 rows on 50 $\mu$m pitch, and covers an area of about 19 mm 
$\times$ 20 mm. It is designed in a 130 nm feature size bulk CMOS process. 
Sensors must be DC coupled to FE-I4 with negative charge collection. The FE-I4 
is very well suited to the AFP requirements: the granularity of cells provides 
a sufficient spatial resolution, the chip is radiation hard enough 
(up to a dose of 3~MGy), 
and the size of the chip is sufficiently 
large that one module can be served by just  one chip. 

The dimensions of the individual cells in the FE-I4 chip are 50 $\mu$m $\times$
250 $\mu$m in the  $x$ and $y$ directions, respectively.
Therefore to achieve the required position resolution in the
$x$-direction of $\sim$ 10 $\mu$m, six layers with sensors are required
(this gives  50/$\sqrt{12}$/$\sqrt{5}\sim 7$ $\mu$m in $x$ and roughly 5 times 
worse in $y$). Offsetting planes alternately to the left and right by one half 
pixel will give a further reduction in resolution of at least 30\%. 
The AFP sensors are expected to be exposed to a dose of 30~kGy
per year at the full LHC luminosity of 10$^{34}$cm$^{-2}$s$^{-1}$.

\subsection{Timing detectors}

A fast timing system that can precisely measure the time difference between 
outgoing scattered protons is a key component of the AFP detector.  The 
time difference is equivalent to a constraint on the event vertex, thus 
the AFP timing detector can be used to reject overlap background by 
establishing that the two scattered protons did not originate from the same 
vertex as the the central system.  The final timing 
system should have the following characteristics~\cite{timing}: 
\begin{itemize}
\item 10 ps or better resolution (which leads to a factor 40 rejection on pile up 
background)
\item Efficiency close to 100\% over the full detector coverage
\item High rate capability (there is a bunch crossing every 25 ns at the nominal LHC)
\item Enough segmentation for multi-proton timing
\item Level trigger capability
\end{itemize}

\begin{figure}
\centering
\includegraphics[width=3.5in]{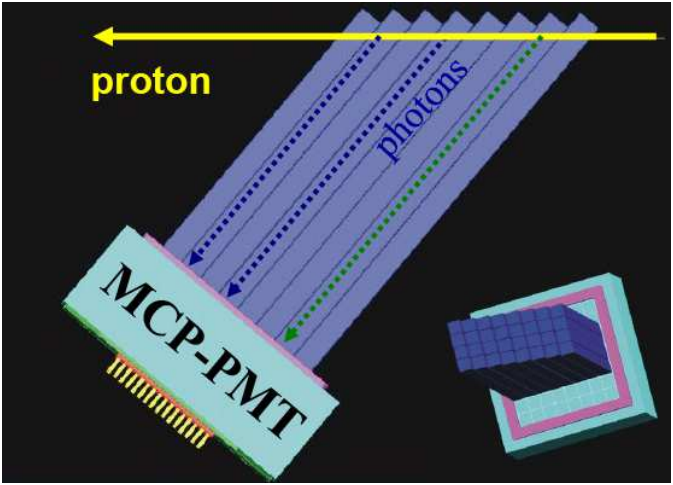}
\vspace*{-0.1in}
  \caption{A schematic diagram of the QUARTIC fast timing detector.
   } \label{fig5_timesys}
\end{figure}

\begin{figure}
\centering
\includegraphics[width=3.in]{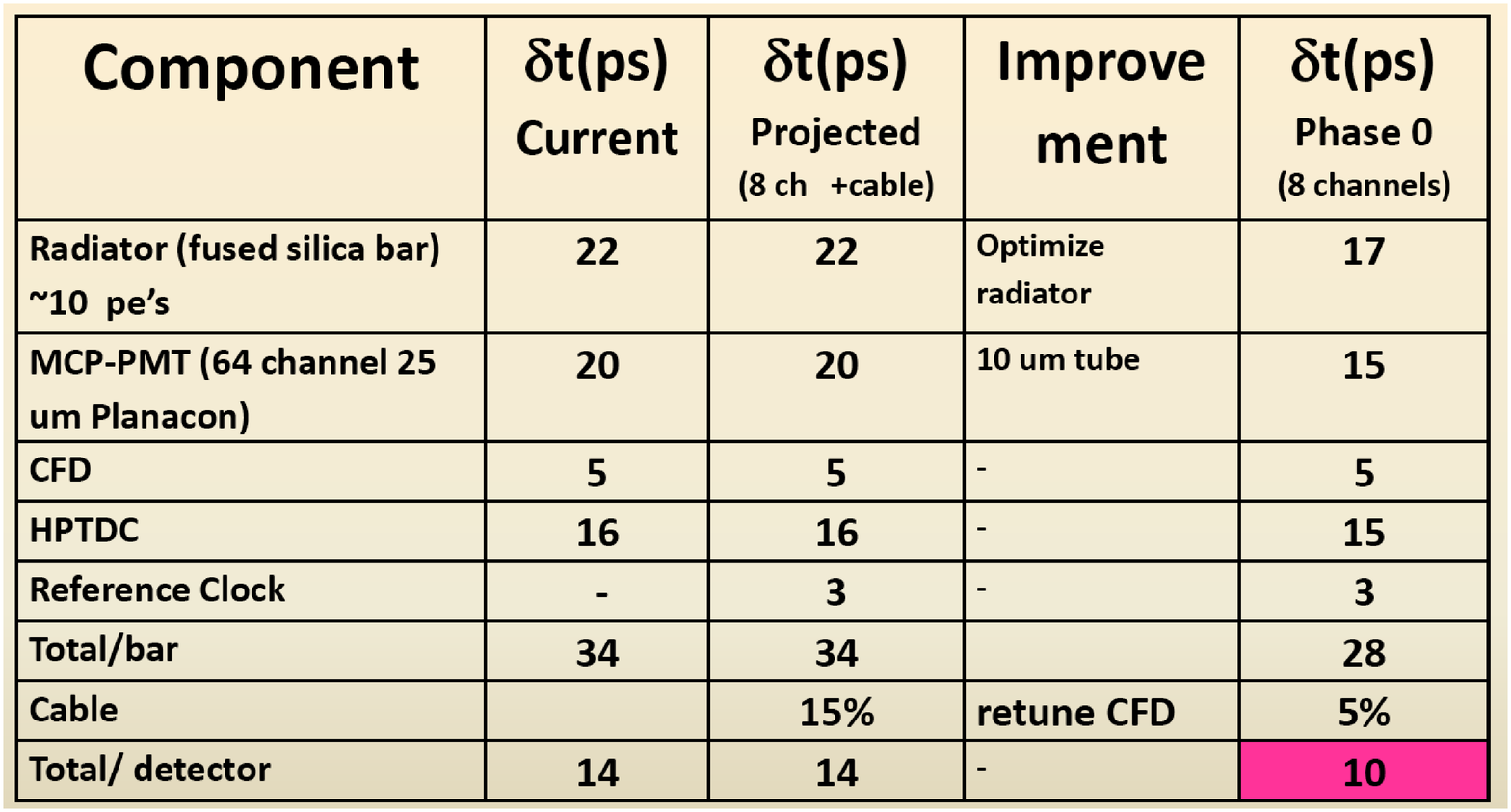}
  \caption{Different components of the timing detector resolution~\cite{timing}.
   } \label{timresol}
\end{figure}

Figure~\ref{fig5_timesys} shows a schematic overview of the first proposed 
timing system, consisting of a quartz-based Cerenkov detector coupled to 
a microchannel plate photomultiplier tube (MCP-PMT), followed by the 
electronic elements  that amplify, measure, and record the time of the event 
along with a stabilized reference clock signal. The
QUARTIC detector consists of  an array of 8$\times$4 fused
silica bars ranging in length from about 8 to 12 cm and oriented at the
average Cerenkov angle.  A proton that is sufficiently deflected from the beam axis will pass 
through a row of eight bars emitting Cerenkov photons providing an overall time resolution that is 
approximately $\sqrt{8}$ times smaller than the single bar resolution of about 30 ps, 
thus approaching the 10 ps resolution goal. Prototype tests have generally been performed on one row 
(8 channels) of 5 mm $\times$ 5 mm pixels, while the initial detector is foreseen to have four rows
to obtain full acceptance out to 20  mm from the beam. The beam tests lead to a time resolution per
bar of the order of 34 ps. The different components of the timing resolution are given in
Fig.~\ref{timresol}. The upgraded design of the timing detector has equal rate pixels, 
and we plan to reduce the the width of detector bins
close to the beam, where the proton density is highest. 

At higher luminosity of the LHC (phase I starting in 2019), higher pixelisation of the timing detector
will be required. For this sake, a R\&D phase concerning timing detector
developments based on Silicon photomultipliers (SiPMs),
avalanche photodiods (APDs), quartz fibers, diamonds has started. In parallel, a new timing readout chip has
been developed in Saclay. It uses waveform sampling methods which give the best possible timing resolution. The
aim of this chip called SAMPIC~\cite{sampic}  (see Fig.~12) is to obtain sub 10 ps timing resolution, 1GHz input bandwidth, no dead
time at the LHC, and data taking at 2 Gigasamples per second. The cost per channel is estimated to be
of the order for \$10 which a considerable improvement to the present cost of a
few \$1000 per channel,
allowing us to use this chip in medical applications such as PET imaging detectors. The holy grail of
imaging 10 picosecond PET detector seems now to be feasible: with a resolution better than 20 ps,
image reconstruction is no longer necessary and real-time image formation becomes
possible.

\section{Acknowledgments}
The author thanks for the financial support from the CAPES/COFECUB exchange program number Ph74412 and from
the local organisers..

\begin{figure}
\centering
\includegraphics[width=3.in]{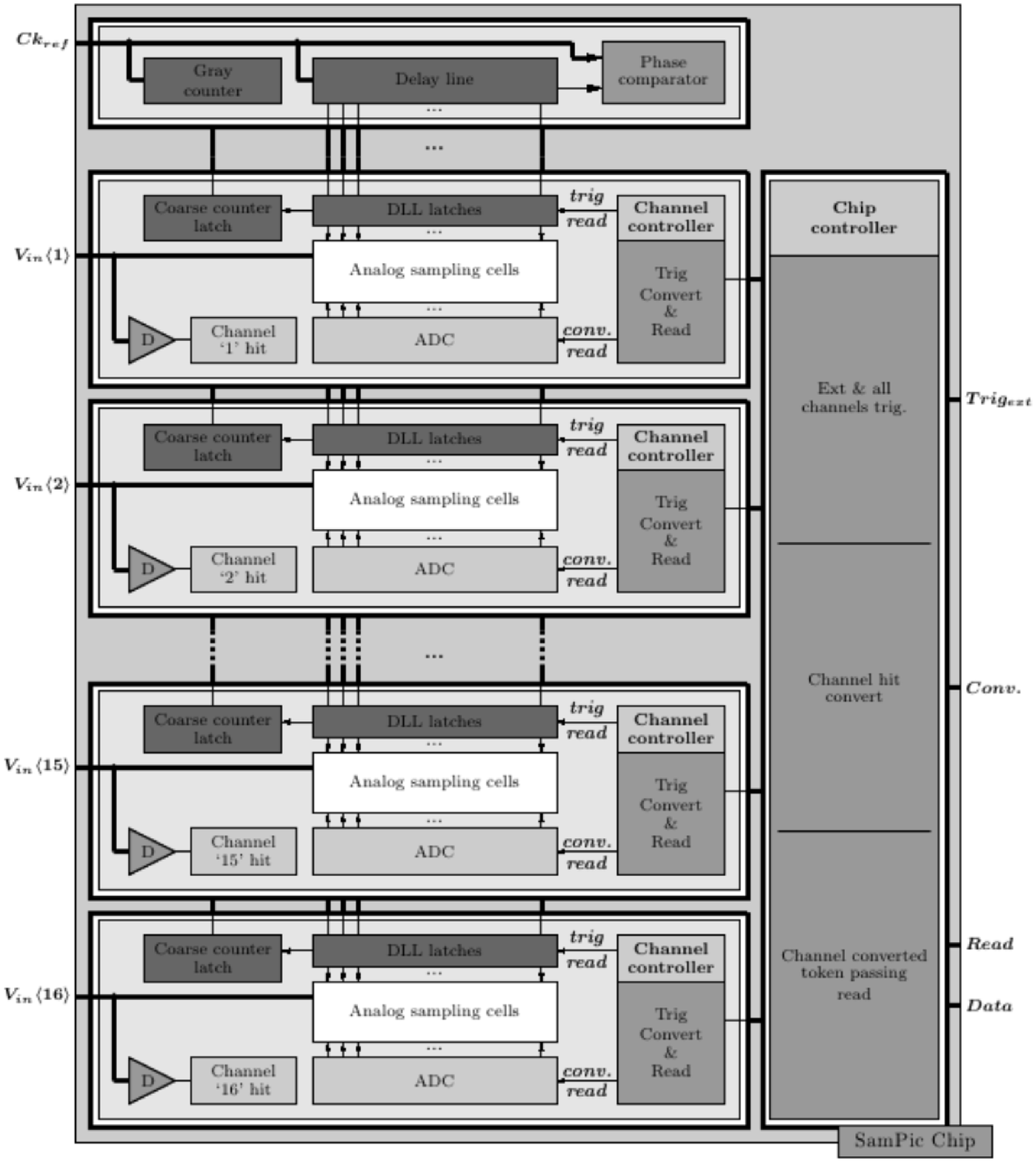}
\caption{Scheme of the sampic chip.}\label{Figa}
\end{figure}

\input{bibliography}

\end{document}
